\documentstyle[11pt,newpasp,twoside,epsf]{article}
\newcommand {\hI} {\ion{H}{I}\,\,}  
  
\newcommand {\ha} {H$\alpha$\,\,}  
\newcommand {\kms} {\,km\,s$^{-1}$\,}  
\newcommand {\M} {\mbox{${\cal M}$}}  
  
\newcommand {\msol} {\M$_\odot$\,}  
\newcommand {\lsol} {\L$_\odot$\,}

\newcommand {\mlb} {(\M/L$_B$)$_\star$\,\,}

\begin{document}
\title{Multiwavelength Rotation Curves to Test Dark Halo Central Shapes}

\author{S\'ebastien Blais-Ouellette \& Claude Carignan} 
\affil{D\'epartement de physique and Observatoire du mont M\'egantic,  
Universit\'e de Montr\'eal, C.P. 6128, Succ. centre ville,  
Montr\'eal, Qu\'ebec, Canada. H3C 3J7 and\\
e--mail: blaisous@astro.umontreal.ca \& carignan@astro.umontreal.ca}
  
\author{Philippe Amram}
\affil{Observatoire Astronomique Marseille-Provence \& Laboratoire d'Astrophysique de Marseille, 2 Place Le Verrier, 13248  Marseille, Cedex 04, France\\  
e--mail: amram@observatoire.cnrs-mrs.fr} 

\begin{abstract}
We use Fabry-Perot \ha spectroscopy, complemented with
published \hI radio synthesis observations to derive high resolution
rotation curves of spiral galaxies. We investigate precisely their
inner mass distribution and compare it to CDM simulations
predictions. Having verified the existence of the so-called core-cusp
problem, we find that the dark halo density inner slope is related
to the galaxy masses. Dwarf galaxies with $V_{max} < 100$ km/s have
halo density inner slope $0 < \gamma < 0.7$ while galaxies with
$V_{max} > 100$ km/s are best fitted by $\gamma \geq 1$.
\end{abstract}  

\section{Introduction}
The dynamical masses of spiral galaxies are known to differ
significantly from their visible masses. The commonly accepted cause
is the existence of a ellipsoidal halo of unseen matter in addition to
the stars and gas. The exact density distribution of these halos have
become an increasingly important issue. On one side, N-body
simulations of the cosmological evolution of the Cold Dark Matter
(CDM) have now reached a sufficient resolution to predict the dark
halo density profiles down to a scale corresponding to the inner parts
of the spiral galaxies (Fukushige et al. 1997; Navarro et al. 1997;
Moore et al. 1998; Ghigna et al. 2000).  They almost always predict
dense cuspy halos. On the other side, observations of dwarf spiral
galaxies show shallow inner rotation curves, compatible with a flat
density core (Blais-Ouellette et al. 2000, hereafter paper II).

The first step in showing the reality of this discrepancy is to
eliminate the known possibilities of systematic observational
biases. Two classes of errors could contribute to underestimate the
velocities, hence the computed density, in the inner parts of spiral
galaxies. The prime culprit in radio observations is the ``beam
smearing'' effect due to the relatively low angular resolution of 21
cm data with sufficient sensitivity to detect \hI in the outer
part of spiral galaxies. Combining the \hI density gradient with the
velocity gradient will lead to underestimate the velocity at a given
radius.

\ha observations always easily reach an angular resolution where any
beam smearing effect can be neglected. A less often commented source
of uncertainties though is found in long slit observations, where most
of \ha data come from. The lack of 2D coverage makes the alignment of
the slit crucial to retrieve the real kinematics of a galaxy. Missing
the kinematical center, which is not always the photometric center
(paper II), or just a few degrees between the slit and the galaxy
position angle will also lead to an underestimation of the velocities.
Inclination estimation, which has to be photometrically determined, is
another major source of uncertainties.  In addition, the presence of a
bar would hardly be noticed and its effect would most probably be
confounded with the rotational kinematics.

In order to hedge ourselves against these biases, we use Fabry-Perot
high resolution \ha spectroscopy combined with published radio
synthesis observations to study the mass distribution of 8 dwarf and
spiral galaxies. In paper II, with a smaller sample, we focused on
modeling the mass distribution using different shape of dark matter
halos. Here, in addition, we address the more precise question of what
inner density slope ($\gamma$) halo can have for a given galaxy type
or mass.

In section 2, we first briefly review the mass modeling
used in the study. Then, in section 3, we look at detailed
mass models of a few galaxies. In section 4, the relation
between $\gamma$ and galaxy mass is discussed followed by
concluding remarks.

\section{Modeling the Mass Distribution}
\label{sec:halos}

To investigate in details the mass distribution of dark matter halos
without a few assumptions on the matter content of spiral galaxies,
one would have to adjust a good dozen of parameters. First, the
luminous matter distribution depends on the disk and bulge
mass-to-light ratios (and their radial gradient), and on the
bulge-to-disk ratio. The \hI contribution have to be corrected for
helium fraction. The dark halos can be non-spherical, in addition to
the five parameters usually used to describe the radial density
distribution. This general distribution function can be expressed as
(see Paper II for details):

\begin{equation}  
\rho(r) = \frac{\rho_0}{(c+(r/r_0)^\gamma) \left(1+(r/r_0)^\alpha\right)^{(\beta-\gamma)/\alpha}}  
\label{eqn:halo}
\end{equation}
where $\rho_0$ and $r_0$ are characteristic density and radius, where
c, included for ease of comparison with other works, can force the
presence of a flat density core, where $\alpha$ and $\gamma$ are
respectively the inverse outer and inner logarithmic slopes, and
$\beta$ the transition parameter.

One could add the distance that is used to calculate the light
distribution, and a central mass which is suspected to exist in most
spirals. 

For some of these constraints, a fixed value is well accepted. The
mild or absent color gradient in spiral galaxies lead to a radially
constant mass-to-light ratio. The helium fraction can be
approximate to its primordial abundance. Distances are hopefully well
constrained by independent means.

Otherwise, essentially three data sets can be used to constraint these
parameters: luminosity distributions (visible an \hI) and velocity
field or, assuming axisymmetry, light profiles and rotation curve. The
visible light profile, is used to determine the bulge to disk
ratio. The other parameters are all to the charge of the rotation
curve. Most of the luminous contributions are heavily constrained by
the most inner parts of the curve leading to a possible degeneracy
between central mass, bulge, inner disk and halo contributions. That
is why, many studies including this one, tend to focus on dwarf
galaxies where bulges are negligible and large central masses excluded
by the rotation curve. Only there, can one put strong limits on the
inner slope of dark halo density distribution. For these galaxies,
luminous matter is dependent only on the disk mass-to-light ratio, and
dark matter from the central density, the core radius, and the three
shape parameters. From the latter, only $\gamma$, the inner slope, has
a significant impact on halo shape at the scale of dwarf galaxies,
$\alpha$ and $\beta$ being very poorly constrained.

In our procedure, the visible light profile (eventually decomposed in
its bulge and disk components) is inverted in a mass profile scaled by
the mass-to-light ratios. The \hI contribution is estimated from its
distribution and multiply by 1.33 to account for helium. The halo
shape is, in general, fixed to a known profile. We choose four
profiles, widely used in the literature, that span, with some
redundancy, through most possibilities. The cuspy profile from Navarro
et al. (1997) is a canonical example of CDM simulation prediction,
the pseudo-isothermal sphere, the profile from Kravtsov et al. (1998),
and Burkert's profile (Burkert et al. 1995) are all cusp-less profile
with different shape parameters.  Rotational velocities from all the
contributions are then added in quadrature and compare to the rotation
curve using a standard $\chi^2$ minimization.

\section{Mass Models}
\label{sec:mm}

\subsection{The Sample}

Beside the dwarf galaxies NGC 3109 and IC 2574 from paper II, the
whole sample includes 3 fairly bulge-less spirals (UGC 2259, NGC 2403,
and NGC 6946), and 3 earlier type spirals (NGC 5055, NGC 2841, and NGC
5985) from paper III. Details are given in Table \ref{tab:sample}.

\begin{table}
\small
\caption{Parameters of the sample. From the main references unless otherwise specified.}
\label{tab:sample}  
\begin{tabular}{l l c c c c c c c c c} \hline\hline
Name & Type & Distance & D$_{25}$ & R$_{HO}$ & $\alpha^{-1}$ & M$_B$ & L$_B$ & \multicolumn{3}{c}{References} \\
&        & Mpc & \arcmin & \arcmin & \arcsec & & $ 10^8$\lsol & \hI & \ha & photom. \\ \tableline
IC 2574 & SABm & 3.0 & 9.76 & 8.63 & 165 & -16.87 & & m & cc & m \\
NGC 3109 & SBm & 1.36$^n$ & 14.4 & 13.3 & 234.9 & & & o & cc & o\\
NGC 5585 & SABd & 6.2 & 5.27 & 3.62 & 46.7 & & & p & gg & p\\
UGC 2259 & SBcd	 & 9.6$^k$ & 2.6 & 1.9 & 28.2  & -16.33 & 5.06 & b & ii & b\\
NGC 2403 & SABcd & 3.2$^h$ & 21.88$$ & 13.0$$ & 134 & -19.50 & 7.9 & c & ii & l \\
NGC 6946 & SABcd & ?  & 11.5 & 7.8 & 1.92 & -21.38 & 530 & d & ii & d \\
NGC 3198 & SBc & 9.15 &&&&&  & c & nn & l \\
NGC 5055 & SAbc	 & 9.2$^k$  & 12.56 & 9.8 & 108.9 & -21.15 & & e & ii & l \\ \tableline
%
\multicolumn{3}{l}{$^{a}${This study}} 	      & \multicolumn{4}{l}{$^{gg}$paperI} & \multicolumn{4}{l}{$^{d}$Carignan et al. 1990} \\
\multicolumn{3}{l}{$^{m}$Martimbeau et al. 1994} & \multicolumn{4}{l}{$^{b}$Carignan et al. 1988} &  \multicolumn{4}{l}{$^{nn}$Corradi et al. 1991} \\
\multicolumn{3}{l}{$^{n}$Musella et al.97}    & \multicolumn{4}{l}{$^{ii}$paperIII} & \multicolumn{3}{l}{$^{e}$Thornley et al. 1997} \\
\multicolumn{3}{l}{$^{cc}$paperII}     &	\multicolumn{4}{l}{$^{h}$Freedman et al. 1990} & \multicolumn{4}{l}{$^{i}$Aaronson et al. 1983} \\
\multicolumn{3}{l}{$^{o}$Jobin et al. 1990}      & \multicolumn{4}{l}{$^{c}$Begeman et al.1987} & \multicolumn{4}{l}{$^{k}$H$_0$ = 75\kms/Mpc}\\
\multicolumn{3}{l}{$^{p}$C\^ot\'e et al. 1991}       & \multicolumn{4}{l}{$^{l}$Kent et al. 1987} & \\
\end{tabular}  
\end{table}

\subsection{The Results}

Complete modeling, using the four models, of the most relevant cases
are presented here. NGC 3109 and IC 2574 (from paper II) are
re-analyze to correct for a numerical problem which underestimated the
central density. They are good examples of CDM incompatible
dwarfs. Are also modeled, NGC 2403, as an example of bulge-less small
spiral compatible with any model, and NGC 5055, as an earlier type
spiral with a bulge. Analysis of the whole sample will be presented in
a future paper (paper IV). Error bars are the quadratic sum of:
half the velocity difference between receding and approaching sides, half the
correction for asymmetric drift, error cause by uncertainty
on inclination, and statistical error($\sigma/\sqrt{N}$).
\begin{figure}
\plotone{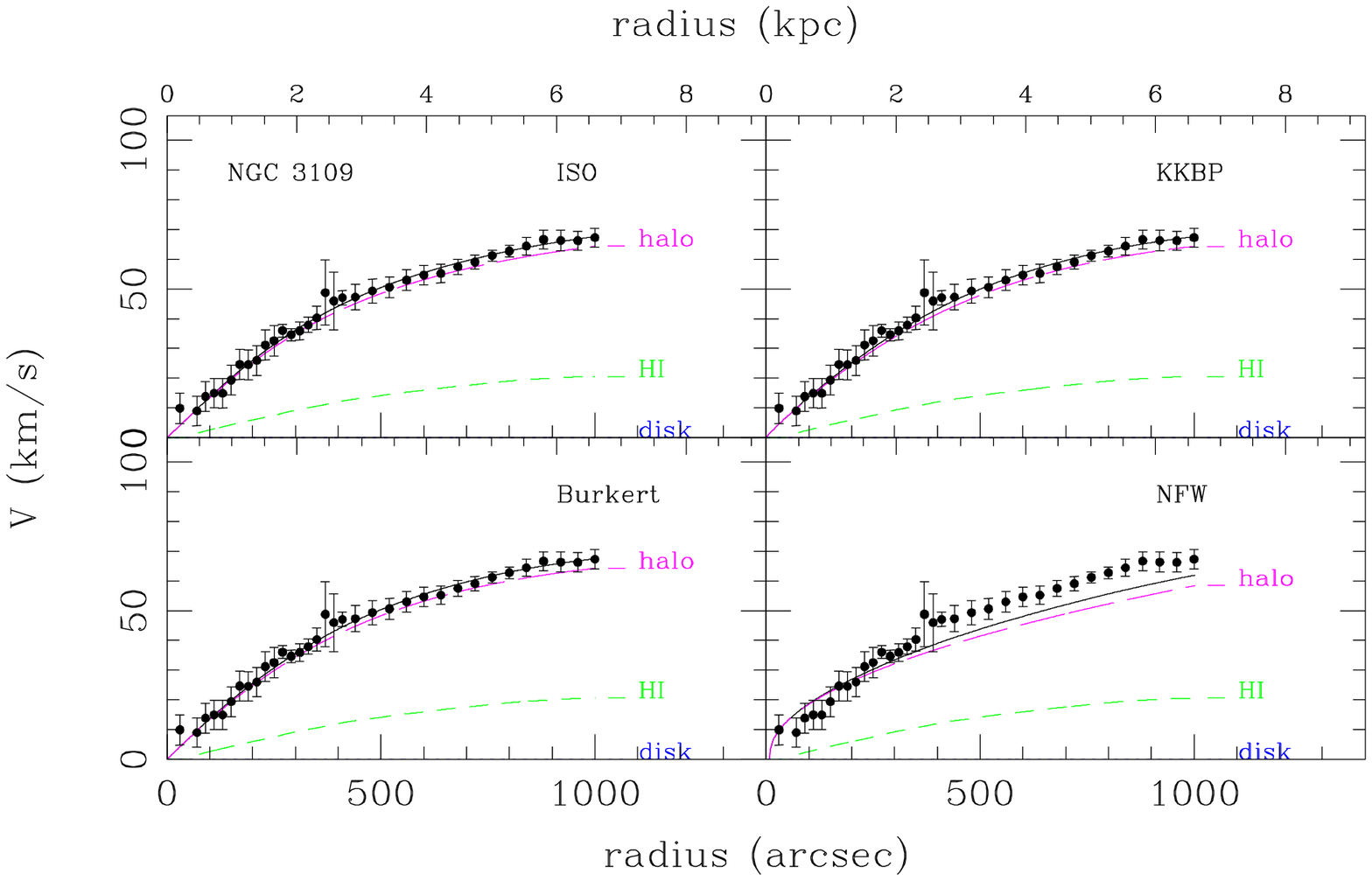} \\
\plotone{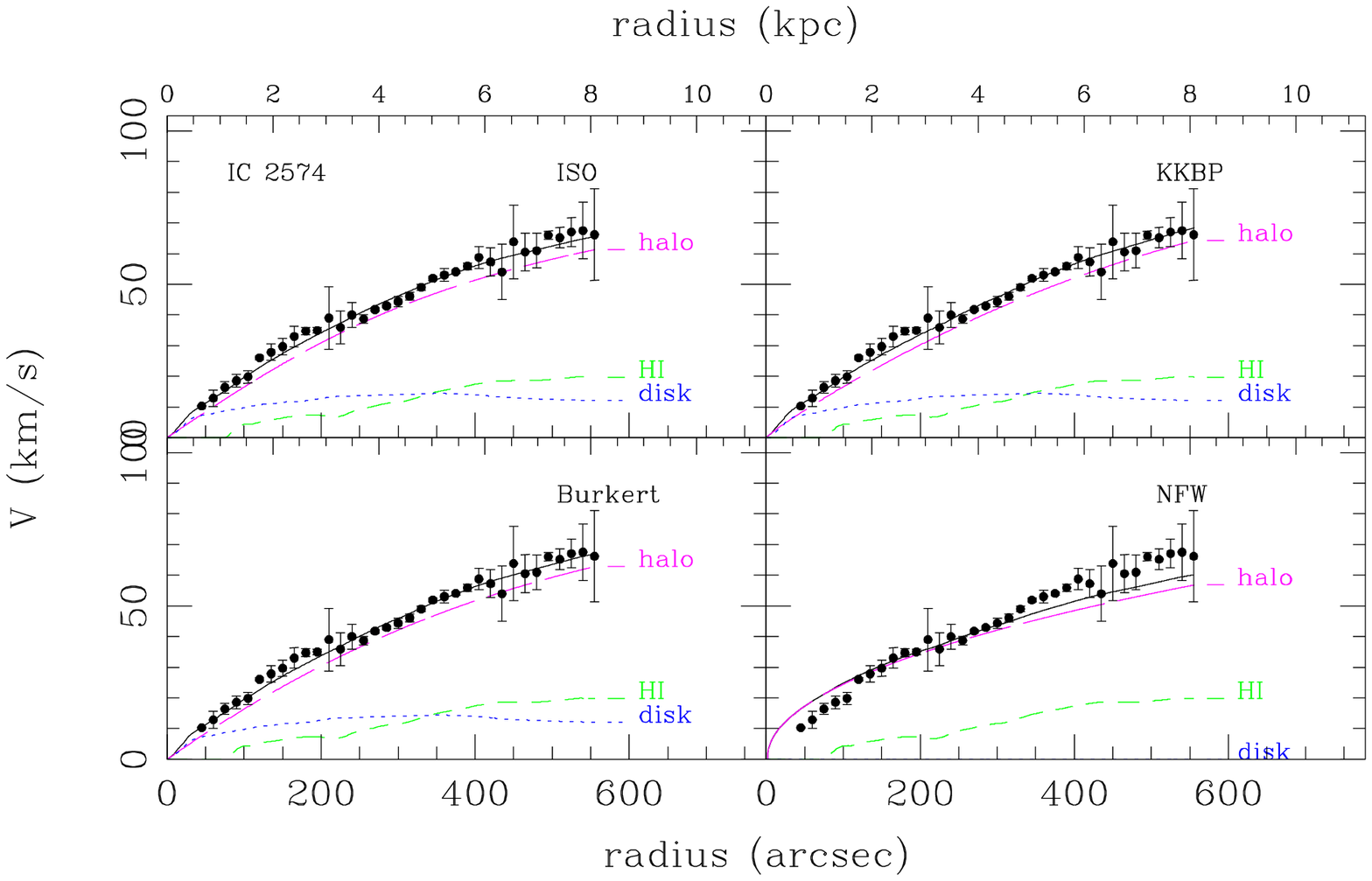} 
\caption{Best fit mass model for NGC 3109 (using the \ha up to
410\arcsec completed by the \hI), and for IC 2574 (\hI only).
\label{fig:mm_3109+2574}}  
\end{figure}

\begin{figure}
\plotone{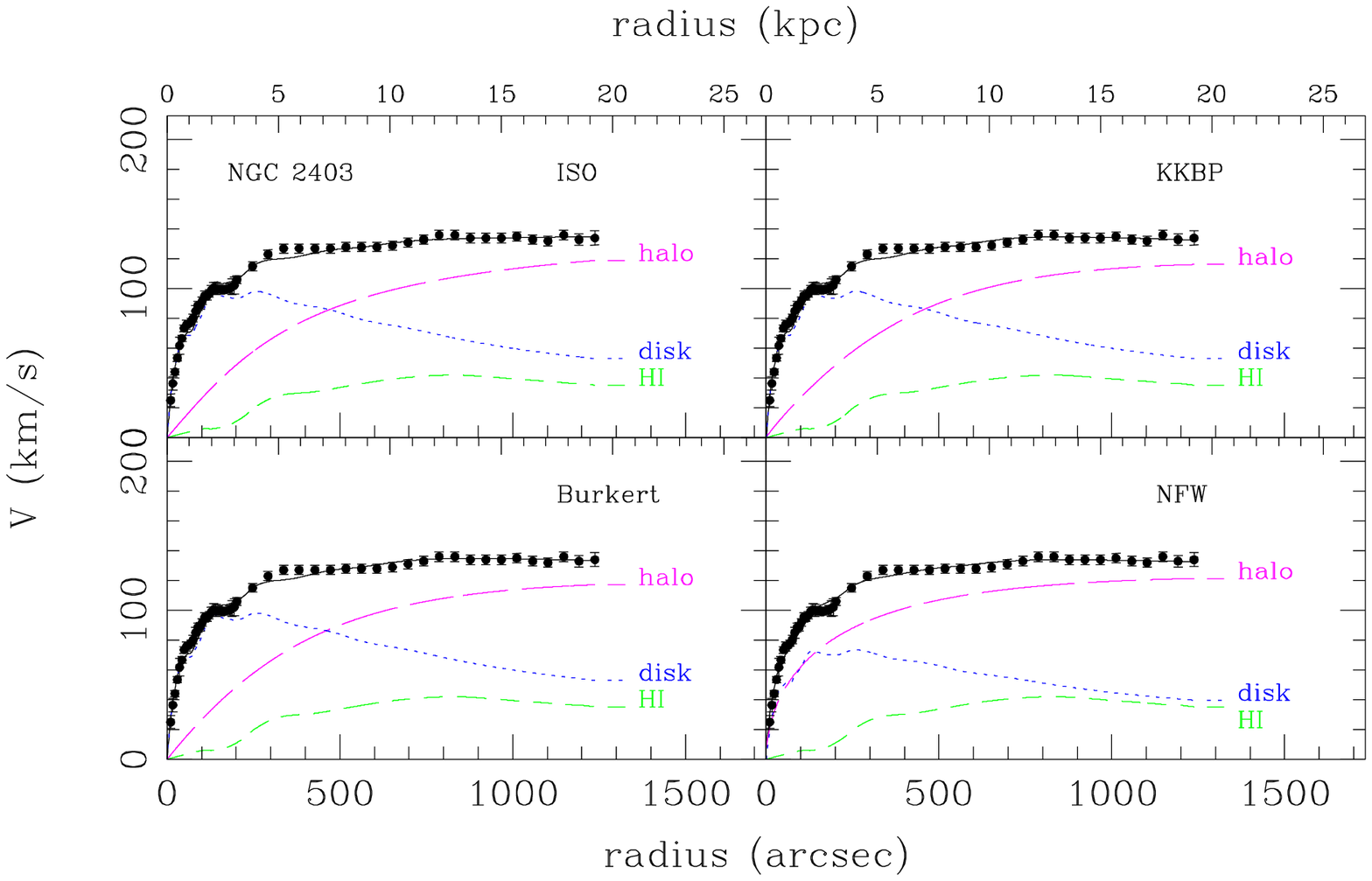}\\ 
\plotone{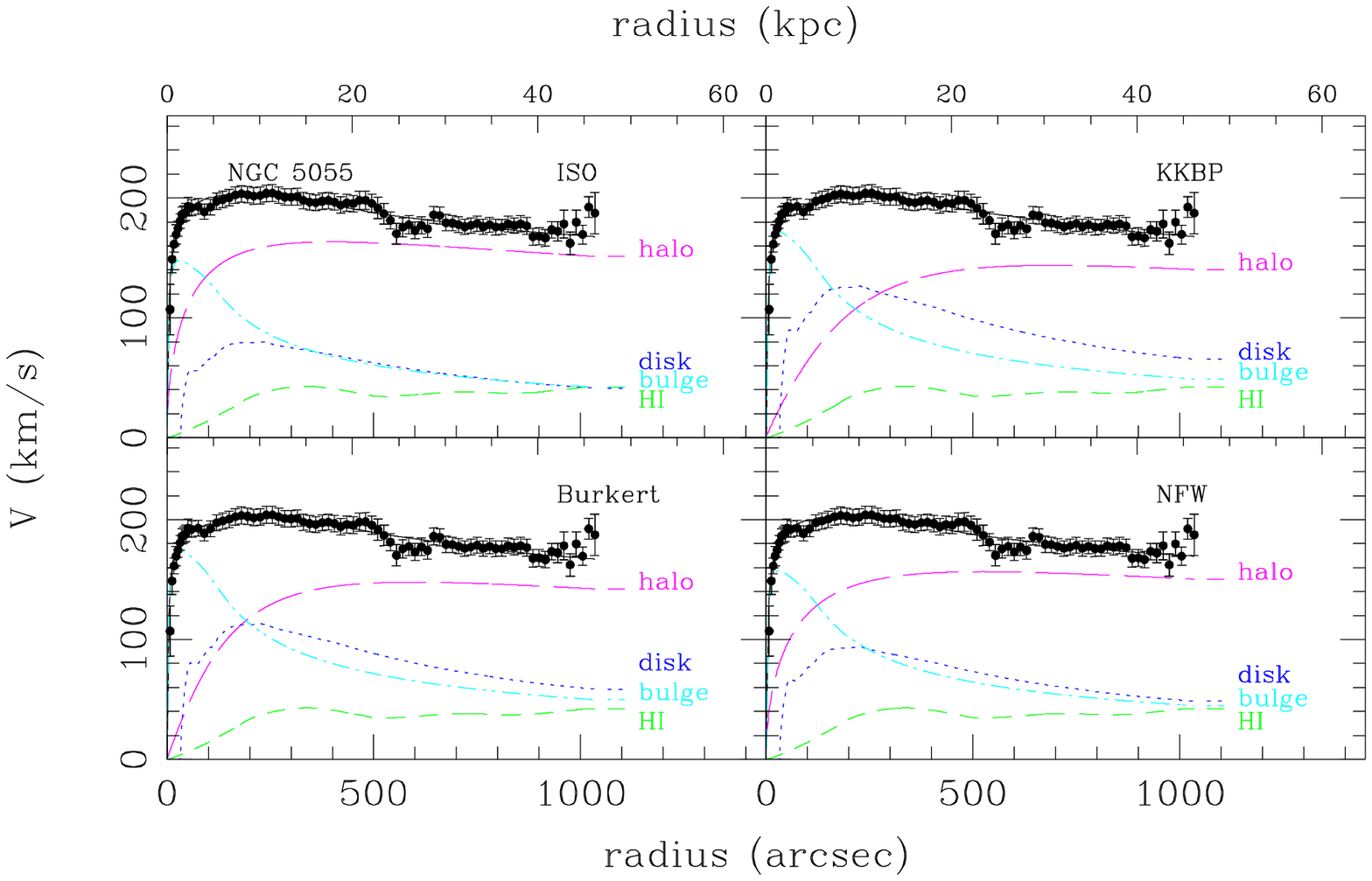}  
\caption{Same as Fig. \ref{fig:mm_3109+2574} for NGC 2403 (using
the \ha up to 345\arcsec), and for NGC 5055 (\ha is used up to 52
\arcsec).
\label{fig:mm_2403+5055_vhiha}}
\end{figure}

\begin{table}[tbh]
\begin{center}
\small
\caption{Parameters of the mass models}
\label{tab:mmIV}  
\begin{tabular}{l l l c c c c c} \hline\hline
Model	& Galaxy & Type  & \mlb bulge & \mlb disk & r$_0$ & $\rho_0$ & $\chi^2$ \\
& & & & & kpc & 10$^{-2}$\msol/pc$^{-2}$ & \\ \hline
\rule[0mm]{0mm}{4mm}ISO 	& NGC 3109 & SBm & no bulge & 0.1 & 2.4 & 2.4 & 0.44 \\
				& IC 2574  & SABm & no bulge & 0.3 & 5.0 & 0.75 & 2.7 \\
				& NGC 2403 & SABcd & no bulge & 2.5 & 4.8 & 1.7 & 1.9 \\
    				& NGC 5055 & SABbc  & 2.3 & 0.8 & 8.2 & 3.4 & 0.55 \\
\rule[0mm]{0mm}{4mm}Burkert 	& NGC 3109 &  & no bulge & 0.4 & 3.0 & 1.8 & 0.22 \\
				& IC 2574  &  & no bulge & 0.3 & 8.9 & 0.8 & 2.5 \\
				& NGC 2403 &   & no bulge & 2.5 & 8.0 & 1.9 & 1.8 \\
    				& NGC 5055 &   & 3.2 & 1.6 & 8.2 & 2.8 & .56 \\
\rule[0mm]{0mm}{4mm}KKBP 	& NGC 3109 &  & no bulge & 0.0 & 4.0 & 1.7 & 0.5 \\
				& IC 2574  &  & no bulge & 0.3 & 10.5 & 0.46 & 2.5 \\
				& NGC 2403 &  & no bulge & 2.5 & 8.5 & 1.2 & 1.8 \\
    				& NGC 5055 &  & 3.1 & 2.0 & 11.2 & 1.0 & .56 \\
\rule[0mm]{0mm}{4mm}NFW 	& NGC 3109 &  & no bulge & 0.0 & 605 & 0.0035 & 9.0 \\
				& IC 2574  &  & no bulge & 0.0 & 295 & 0.005 & 32 \\
				& NGC 2403 &  & no bulge & 1.4 & 11.8 & 0.9 & 1.4 \\
    				& NGC 5055 &  & 2.6 & 1.1 & 10.8 & 1.8 & .52 \\ \hline
\multicolumn{4}{l}{r$_0$: core radius of the dark halo}\\
\multicolumn{4}{l}{$\rho_0$: central density of the dark halo}\\
\end{tabular}
\end{center}  
\end{table}

It is rather clear that NFW profile is incompatible with the two dwarf
galaxies. These results are in line with most similar studies using
\hI (e.g. C\^ot\'e et al. 2000) or long slit observations (e.g.  Swaters
et al. 2000, de Blok et al. 2001) but without the related
uncertainties (van den Bosch et al. 2000). It has to be noticed that
higher resolution N-body simulations tend to give even steeper inner
density slopes. The present discrepancy is therefore genuine and
probably involve subtle phenomena or new physics.

Mass-to-light ratios tend to be unrealistically low for the smallest
galaxies. Halo contributions are therefore an upper limit
and are in fact probably shallower.

\section{Halo Density Gradient}
\label{sec:gamma}

\begin{figure}[tbh]
\begin{center}
\plotfiddle{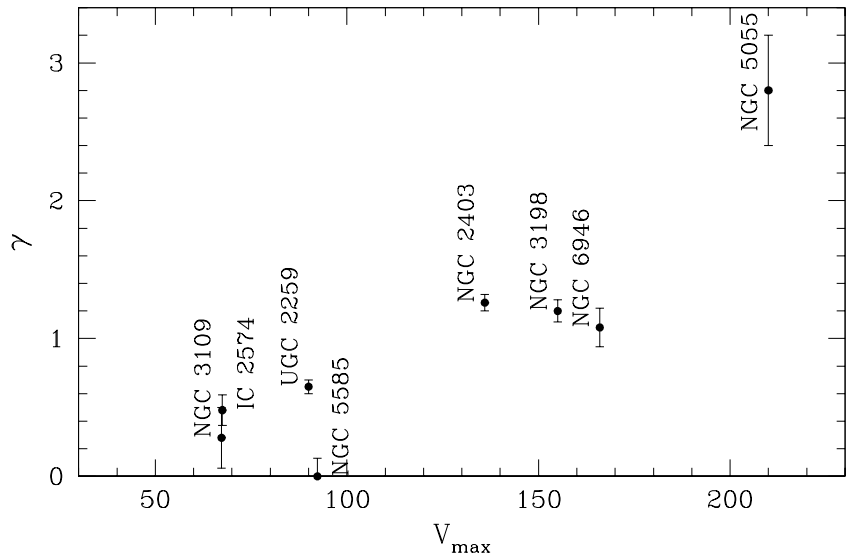}{7cm}{0}{60}{60}{-180}{-210}
\caption{Inner logarithmic slope ($\gamma$) of the CDM halos.  Note
that N-body simulations give $\gamma \geq 1$.
\label{fig:gammax}}
\end{center}
\end{figure}

The range of possible observational biases that could explain the
core-cusp problem is now significantly reduced, rather close to
nil. The question of the physical cause of this shallow inner density
distribution have been addressed many times leading to proposals ranging
from multi-component dark matter (Burkert \& Silk 1997) to
self-interacting dark matter (Spergel \& Steinhardt) and baryonic
feedback processes. These scenarios are though to behave differently
under different gravitational potentials. For example, feedback
processes can hardly have an impact in galaxies with very deep
potential. Plotting $\gamma$, the density profile inner slope, against
the asymptotic rotational velocity would precisely show the evolution
of this behavior. Figure \ref{fig:gammax} plots $\gamma$ against the
observe maximum velocity which, for NGC 3109 and IC 2574, is lower
than the true asymptotic velocity. 

The steepness of NGC 5055 rotation curve leads to an important
degeneracy between $\gamma$, r$_0$ and $rho_0$. It is therefore not
clear from the plot if there is a canonic value of $\gamma$ around 1.2
that breaks down below 100 km/s or if $\gamma$ is intrinsically
increasing in massive galaxies.

\section{Conclusion}

We have modeled 8 dwarf and spiral galaxies using cusp-less and cuspy
halos. The latter, predicted by CDM N-body simulations are clearly
incompatible with dwarf galaxy kinematics. More precisely, galaxies
with less than 100 km/s of maximum rotation velocity have a inner
density logarithmic slope ($\gamma$) of less than .7 as opposed to
$\gamma \ge 1$ predicted by the N-body simulations.

Presently, $\alpha$, the outer density slope is poorly constrained by
rotation curves. One need to reach a radius where the luminous disk
contribution is negligible while the rotation curve is flat or
decreasing (Carignan \& Purton 1998). In our sample, UGC 2259 is a
good candidate.

Future studies should in part extend to massive bulge-less galaxies to
see if $\gamma$ is intrinsically rising with galaxy mass or if it is
constant, with a break down at low mass. The GHASP survey (Garrido \&
Amram 2002) should allow this kind of study.


\begin{references}
\small

\reference {}Aaronson, M. \& Mould, J. 1983, \apj, 265, 1
\reference {}Begeman,K 1987, PhD Thesis, Rijksuniversiteit Groningen
\reference {}Blais-Ouellette, S., Carignan, C. \& Amram, P., \& C\^ot\'e, S. 1999 \aj, 118, 2123 
\reference {}Blais-Ouellette, S., Amram, P. \& Carignan, C. 2001 \aj, 121, 1952
\reference {}Burkert, A. 1995, \apj, 447, L25
\reference {}Carignan, C. \& Freeman, K.C. 1988, \apj, 332, L33
\reference {}Carignan, C., Charbonneau, P., Boulanger, F. \& Viallefond, F. 1990, \aap, 234, 43
\reference {}Carignan, C. \& Purton, C. 1998, \apj, 506, 125
\reference {}C{\^ o}t{\' e}, S., Carignan, C., \& Freeman, K.~C. 2000, \aj, 120, 3027
\reference {}C\^ot\'e, S., Carignan, C. \& Sancisi, R. 1991, \aj, 102, 904
\reference {}Corradi, R.L.M., Boulesteix, J., Bosma, A., Capaccioli, M., Amram, P. \& Marcelin, M. 1991, \aap, 244, 27
\reference {}de Blok, W.~J.~G., McGaugh, S.~S., Bosma, A. \& Rubin, V.~C. 2001, \apj, 552, L23
\reference {}Freedman, W. L. 1990, \apj, 335, L35
\reference {}Fukushige \& T., Makino, J. 1997, \apj, 477, L9 
\reference {}Garrido, O., Amram, P., This conference
\reference {}Ghigna, S., Moore, B., Governato, F. ,Lake, G. ,Quinn, T. \& Stadel, J. 2000, 544, 616
\reference {}Jobin, M. \& Carignan, C. 1990, \aj, 100, 648
\reference {}Kent, S. M. 1984, \apjs, 56, 105
\reference {}Kravtsov, A. V., Klypin, A. A., Bullock, J. S. \& Primack, J. R. 2000,\apj, 502, 48
\reference {}Martimbeau, N., Carignan, C. \& Roy, J.-R. 1994, \aj, 107, 543
\reference {}Moore, B., Governato, F., Quinn, T., Stadel, J. \& Lake, G. 1998, \apj, 498, L5 
\reference {}Musella, I., Piotto, G. \& Capaccioli, M. 1997, \aj, 114, 976
\reference {}Navarro, J.F., Frenk C.S. \& White, S.D.M. 1997, \apj, 490, 493 
\reference {}Swaters, R. A., Madore, B. F. \& Trewhella, M. 2000, \apj, 531, L107
\reference {}Thornley, M. D. \& Mundy, L. G. 1997, \apj, 484, 202
\reference {}van den Bosch, F. C., Robertson, B. E., Dalcanton, J. J. \& de Blok, W. J. G. 2000, \aj, 119, 1579


\end{references}
\end{document}